\documentclass[iop]{emulateapj}
\usepackage{xspace,graphics,graphicx}

\newcommand\teff{\ensuremath{T_{\rm eff}}\xspace}
\newcommand\logg{\ensuremath{\log g}\xspace}

\newcommand\dqv{SDSS J1036+6522\xspace}

\slugcomment{Accepted for publication in the Astrophysical Journal} 
\shorttitle{The DQV SDSS J1036+6522}
\shortauthors{Williams et al.}

\begin{document}
\title{Photometric variability in a warm, strongly magnetic DQ white dwarf, SDSS J103655.39$+$652252.2}

\author{Kurtis A.~ Williams}
\affil{Department of Physics \& Astronomy \\ Texas A\&M
  University-Commerce \\ P.O.~Box 3011, Commerce, TX, 75429, USA}
\email{Kurtis.Williams@tamuc.edu} 

\author{D.~E.~ Winget}
\author{M.~H.~ Montgomery} 
\affil{Department of Astronomy \\ University
  of Texas \\ 1 University Station C1400, Austin, TX, 78712, USA}
\author{Patrick Dufour}
\affil{D\'epartement de Physique \\ Universit\'e de Montr\'eal \\
  C.P. 6128, Succ. Centre-Ville, Montr\'eal, QC H3C 3J7, Canada}
\author{S.~O.~Kepler} 
\affil{Departamento de Astronomia \\ Universidade
  Federal do Rio Grande do Sul \\ Av. Bento Gon\c{c}alves 9500 Porto
  Alegre 91501-970, RS, Brazil}
\author{J.~J.~ Hermes}
\author{Ross E.~ Falcon}
\author{K.~I.~ Winget}
\affil{Department of Astronomy \\ University
  of Texas \\ 1 University Station C1400, Austin, TX 78712, USA}
\author{Michael Bolte}
\affil{UCO/Lick Observatory \\ University of California\\ 1156 High St., Santa Cruz, CA, USA, 95064}
\author{Kate H.~R. Rubin}
\affil{Max-Planck-Institut f\"ur Astronomie \\ K\"onigstuhl 17, 69117 Heidelberg, Germany}
\and
\author{James Liebert}
\email{jamesliebert@gmail.com}

\begin{abstract}
  We present the discovery of photometric variability in the DQ white
  dwarf SDSS J103655.39+652252.2 (\dqv).  Time-series photometry reveals a coherent monoperiodic modulation at a period of
  $1115.64751(67)$ s with an amplitude of $0.442\% \pm 0.024\%$; no other
  periodic modulations are observed with amplitudes $\gtrsim 0.13\%$. The
  period, amplitude, and phase of this modulation are constant within
  errors over 16 months.  The spectrum of \dqv shows
  magnetic splitting of carbon lines, and we use Paschen-Back
  formalism to develop a grid of model atmospheres for mixed carbon
  and helium atmospheres.  Our models, while reliant on several
  simplistic assumptions, nevertheless match the major spectral and
  photometric properties of the star with a self-consistent set of
  parameters: $\teff\approx 15,500$K, $\logg\approx 9$,
  $\log(\mathrm{C}/\mathrm{He})=-1.0$, and a mean magnetic field
  strength of $3.0\pm 0.2$ MG. The temperature and abundances strongly suggest that \dqv is a transition object between the hot, carbon-dominated DQs and the cool, He-dominated DQs. The variability of \dqv has characteristics similar to
  those of the variable hot carbon-atmosphere white dwarfs
  (DQVs), however, its temperature is significantly cooler. 
  The pulse profile of \dqv is nearly 
  sinusoidal, in contrast with the significantly asymmetric pulse
  shapes of the known magnetic DQVs. 
  If the variability in \dqv is due to the same mechanism as other DQVs, then the pulse shape is not a definitive diagnostic on the
  absence of a strong magnetic field in DQVs.  It remains
  unclear whether the root cause of the variability in \dqv and the
  other hot DQVs is the same.
  
\end{abstract}

\section{Introduction}

White dwarfs (WDs), the final stage of stellar evolution for the vast
majority of stars, exhibit a menagerie of spectral characteristics.
This is somewhat surprising at first, as the extreme surface gravities
of WDs should chemically stratify their atmospheres on astronomically
short timescales \citep[e.g.][]{1992ApJS...82..505D,2003ApJ...596..477Z}.  The origins of the spectral species of WDs are
thought to be either a result of the final stages of stellar
evolution, such as a late thermal pulse exhausting the residual
hydrogen in a star \citep[e.g.][]{1983ApJ...264..605I}, or as a result of physical processes along the WD
cooling track, such as convective mixing and accretion \citep[e.g.][]{1976A&A....52..415K,1993ApJS...84...73D,2008ApJ...672.1144T}.  An
understanding of these origins is crucial for many areas of
astrophysical interest, such as the cooling rates (and therefore
age-dating) of WD populations \citep[e.g.][]{1987ApJ...315L..77W,2004ApJS..155..551H,2005ApJ...622..565V}, the fates of planetary systems \citep[e.g.][]{2003ApJ...584L..91J,2007ApJ...671..872Z,2009ApJ...694..805F}, and
physical processes during the post main-sequence evolution of stars \citep[e.g.][]{1990ARA&A..28..103W,1999ApJ...520..680H,2001PASP..113..409F,2008ARA&A..46..157W}.

Using spectral data from the Sloan Digital Sky Survey (SDSS),
\citet{2003AJ....126.2521L} identified eighteen WDs of spectral class
DQ whose spectra exhibit lines of neutral and/or ionized atomic carbon,
while the majority of DQ WDs have spectral features indicative of
molecular carbon, such as Swan bands.  A few years later, another DQ
WD with features of atomic carbon was discovered in the field of the
young open star cluster \object{Messier 35}
\citep{2006ApJ...643L.127W}. Both papers suggest that these WDs had
helium-dominated atmospheres with traces of carbon sufficient to mask
the helium spectral features.  Shortly thereafter,
\citet{2007Natur.450..522D,2008ApJ...683..978D} recognized that,
instead, these ``hot DQ" WDs ($\teff \approx 18,000-24,000$ K) have carbon-dominated atmospheres, with carbon
abundances $\log({\rm C/He}) \gtrsim 1$.

The origin of these carbon-dominated atmospheres in a relatively small
temperature range is difficult to explain.
\citet{2008ApJ...683..978D} propose a spectral evolution scenario,
expanded upon by evolutionary calculations of
\citet{2009ApJ...693L..23A}.  In this scenario, a hot WD undergoes a
vigorous very late thermal pulse that reduces the residual He content
of the star to $\sim 10^{-8}-10^{-7} M_{\rm WD}$.  Immediately
following the very late thermal pulse, the WD would look like the very
hot, carbon/oxygen-rich WD \objectname{H1504+65}
\citep[e.g.][]{2004A&A...421.1169W}.  The residual helium would then
rise to the surface of the WD, causing it to evolve to spectral type
DB (helium-atmosphere), until a convective zone mixes this helium
veneer back into the carbon-rich lower atmosphere at roughly the
temperature where the hot DQs appear.

Alternatively, hot DQ WDs could be remnants of the so-called
``super-AGB" stars, stars massive enough to ignite carbon,
resulting in an oxygen-neon WD surrounded by a layer of carbon and
oxygen; \citet{1997ApJ...485..765G} predict that such a star
could have no residual hydrogen and little, if any, residual helium.
This latter scenario requires that all hot DQ WDs should be massive
($\gtrsim 1M\sun$), while the former scenario allows for lower-mass
hot DQ stars.

WD asteroseismology is a well-established method for studying the
parameters and interior structures of WDs, provided non-radial
pulsations are present and can be detected.  Simple arguments involving the thermal
timescale at the base of the convection zone led \citet{2008ApJ...678L..51M} to suspect that
some hot DQs might be unstable to non-radial pulsations, and
observations inspired by this possibility led to the discovery of
variations in the hot DQ \objectname{SDSS J142625.70+565218.4} (hereafter SDSS J1426$+$5652).  Nearly
simultaneous with the publication of this discovery,
\citet{2008A&A...483L...1F} published a prediction based on detailed
nonadiabatic calculations that hot DQs may be unstable to nonradial
pulsations if sufficient helium is present in the atmosphere.
Subsequently, \citet{2009A&A...506..835C} published nonadiabatic
pulsation analysis of hot DQs based on their hot DQ evolutionary
models.  Crucially, all three of these pulsational stability analyses
present significantly different predictions of the location of the hot DQ instability
strip in \teff ~and \logg ~space.

Observations of variable hot DQs have uncovered additional mysteries
about these stars.  The folded pulse shape of SDSS J1426$+$5752
differs significantly from those of pulsating DA and DB stars
\citep{2008ApJ...678L..51M}, and the star's spectrum shows Zeeman
splitting indicative of a strong magnetic field \citep[$\sim 1.2$
MG,][]{2008ApJ...683L.167D}.  Following the discovery of variability
in SDSS J1426$+$5752, \citet{2008ApJ...688L..95B} detected variability
in two additional hot DQ stars, \objectname{SDSS J220029.08-073121.5}
(hereafter SDSS J2200$-$0741) and \objectname{SDSS
  J234843.30-094245.3} (hereafter SDSS J2348$-$0942).  They claim both
variable hot DQs have pulse shapes similar to SDSS J1426$+$5752.
Additional observations of these two stars by
\citet{2009ApJ...703..240D} agree with \citet{2008ApJ...688L..95B}
that SDSS J2200$-$0741 has a pulse shape similar to SDSS J1426$+$5752,
but they find that SDSS J2348$-$0942 has a pulse shape more typical of
other classes of pulsating WDs.  This, combined with suggestions of
Zeeman splitting in the spectrum of SDSS J2200$-$0741, leads
\citet{2009ApJ...703..240D} to suggest that SDSS J2200$-$0741 also has
a strong magnetic field, while SDSS J2348$-$0942 does not.

\citet{Dunlap2010} announced the discovery of the
DQV \objectname{SDSS J133710.19$-$002643.5} (hereafter SDSS
J1337$-$0026).  This star has low-amplitude ($\sim 0.3\%$) photometric
variations and has a strongly non-sinusoidal pulse shape \citep{Dunlap2010}.  While their low-resolution,  high signal-to-noise spectrum shows no evidence for Zeeman splitting, a higher-resolution spectrum shows clear Zeeman splitting (P.~Dufour, in preparation).  Could this star have a magnetic field
strong enough to distort the pulse shape (by whatever physical mechanism),
yet be weak enough to not cause noticeable Zeeman splitting in
low-resolution spectra?  Or is there no relation between magnetism and
the pulse shape? 

More recently, \citet{Dufour2011} discovered variablility in yet
another hot DQ, \objectname{WD 1150$+$012} (SDSS J1153$+$0056), based
on variable photon flux rates in the FUV spectrum obtained by the
Cosmic Origins Spectrograph on the Hubble Space Telescope.  These
observations were not long enough to establish a pulse shape for this
object.

Most of these studies have been conducted under the presumption that the variability in DQVs is due to nonradial pulsations.  In addition to initial models suggesting that DQVs may be unstable to such pulsations \citep{2008ApJ...678L..51M,2008A&A...483L...1F}, claims have been made of the detection of additional modes of variability not harmonically related to the dominant mode \citep{2009ApJ...703..240D,2009ApJ...702.1593G}.  If confirmed, these additional modes would be evidence in favor of pulsations.  However, other mechanisms for variability have been postulated, including the musings of \citet{2008ApJ...678L..51M} of accretion of carbon-rich material in an AM CVn-like system and the hypothesis floated by \citet{2008ApJ...683L.167D} that the oscillations could be analogous to those observed in rapidly-oscillating Ap stars.  Other potential causes could include short-period binary orbits or rapid rotation of these white dwarfs.  We discuss these possibilities more thoroughly in Section~\ref{sec.discuss}.

The growing number of variable hot DQ WDs should help to address many of
the conundrums surrounding these objects.  In particular, mapping the
hot DQ instability strip can test the three independent predictions of
pulsational properties and stability.  Additional pulse profiles can
test whether the pulse shapes of variable hot DQs are influenced by
the presence or absence of a magnetic field.  Finally, additional
variables will help to define further the emerging characteristics of
this enigmatic class of variable star.

In this paper, we announce the discovery of variability in another
DQ star, \objectname{SDSS J103655.39+652252.2}\footnote{a.k.a. \objectname{SDSS J103655.38+652252.1} in \citet{2003AJ....126.2521L} and \objectname{SDSS J103655.38+652252.0} in \citet{2006ApJS..167...40E}}, hereafter \dqv.
\dqv was identified by \citet{2003AJ....126.2521L} as a WD
with absorption lines of atomic carbon.  \citet{2003ApJ...595.1101S}
identify three sets of Zeeman-split \ion{C}{1} lines in the star's
spectrum; the splitting implies a mean field strength of 2.9 MG, or a
dipole field strength of $\sim 4$ MG.  Our spectral analysis (Section~\ref{sec.spec}) finds that \dqv is
significantly cooler than any of the other known DQVs.  In Section~\ref{sec.discuss} we discuss the
potential implications of this discovery for the definition and characteristics of DQV WDs.

\section{Time-Series Photometry and Analysis}

Positional and photometric parameters for \dqv are taken from the
Sloan Digital Sky Survey (SDSS) Data Release 8
\citep{2009ApJS..182..543A}. \dqv is located at RA$=10^{\mathrm h}
36^{\mathrm m} 55\fs39$, Dec$=+65\degr 22\arcmin 52\farcs 2$ (J2000).
The star's photometric properties via SDSS point-spread function
photometry are $g=18.514\pm 0.024$, $u-g=-0.217\pm 0.032$, and
$g-r=-0.325\pm 0.031$.  The star has a significant proper motion,
$\mu_{\alpha}=26\pm 1$ mas yr$^{-1}$; $\mu_{\delta}=4\pm 6$ mas
yr$^{-1}$ \citep{2003AJ....125..984M}.

\dqv was included in the \citet{2003AJ....126.2521L} catalog of SDSS WDs
with spectral features of atomic carbon.  At the time of our first
observing run, the spectral parameters \teff and \logg were
unknown.  The star's selection was motivated by its very blue colors,
which suggested it might be of similar temperature to the other known
DQVs.

Time-series photometry of \dqv was obtained over several nights in
late 2008, early 2009, and early 2010 with the Argos high-speed
photometer on the McDonald Observatory 2.1m Otto Struve Telescope
\citep{2004ApJ...605..846N}.  All observations were taken through a 1 mm
Schott glass BG40 filter.  A log of observations is presented in Table
\ref{tab.obslog}.

\begin{deluxetable*}{lccccc}
\tablecolumns{6}
\tablewidth{0pt}
\tablecaption{Observation Log\label{tab.obslog}}
\tablehead{\colhead{UT Date} & \colhead{Run} & \colhead{Start} &
  \colhead{Exp. Time} & \colhead{Run length} & \colhead{Seeing} \\ & \colhead{Name} & \colhead{(BJD$_{\rm TDB}$)} &
  \colhead{(s)} & \colhead{(s)} & \colhead{(arcsec)}}
\startdata
2008 Dec 29 & A1807 & 2454829.88044726 & 10 & 11000 & 1.4 \\
2008 Dec 31 & A1811 & 2454831.79331520 & 30 & \phantom{1}7560 & 1.8 \\
2009 Jan 02 & A1816 & 2454833.80946591 & 30 & 20430 & 2.6 \\
2009 Jan 03 & A1820 & 2454834.91942706 & 30 & 11130 & 1.6 \\
2009 Jan 04 & A1822 & 2454835.78627879 & 30 & 22530 & 1.3 \\
2009 Apr 21 & A1858 & 2454942.64579972 & 15 & 17400 & 1.1 \\
2009 Apr 22 & A1861 & 2454943.61438557 & 15 & 22455 & 1.5 \\
2009 Apr 30 & A1871 & 2454951.59850661 & 15 & 15300 & 1.5 \\
2010 Jan 16 & A2059 & 2455212.82071991 & 15 & 19035 & 2.3 \\
2010 Feb 13 & A2076 & 2455240.69756942 & 15 & 26595 & 2.2 \\
2010 Feb 14 & A2080 & 2455241.87175010 & 15 & 13758 & 1.3 \\
2010 Apr 12 & A2117 & 2455298.63301727 & 15 & 17400 & 1.9 \\
\enddata
\tablecomments{Start time is midpoint of first exposure}
\end{deluxetable*}

The photometric data were reduced with the methods and pipeline
described in \citet{2005ApJ...625..966M} and improved upon in \citet{2008ApJ...676..573M}. Weighted aperture photometry was performed using a variety of aperture radii; the final aperture radius was chosen based on the lowest noise level in the resulting discrete Fourier transform and was typically close to the observed full-width at half maximum (FWHM).  The light curve was divided by a weighted combination of  one to three nearby reference stars chosen to be as similar in color as possible to the target; the precise number of reference stars varied from observing run to observing run; fainter comparison stars were excluded if their light curves were of low signal-to-noise due to less-than-ideal observing conditions.  Differential atmospheric extinction was removed by fitting a second-order polynomial to the light curve, and sections of the light curve were excluded from further analysis if clearly affected by cloud, cosmic ray hits, obscuration by the telescope dome due to inattentive observers, etc.  The data are corrected for the motion of the Earth around the Solar System barycenter using the method of \citet{1980A&AS...41....1S}, and all UTC leap-seconds up to the date of observation are accounted for.  No attempt was made at absolute photometric calibration.

\begin{figure}
\begin{center}
\includegraphics[angle=270,
width=0.45\textwidth]{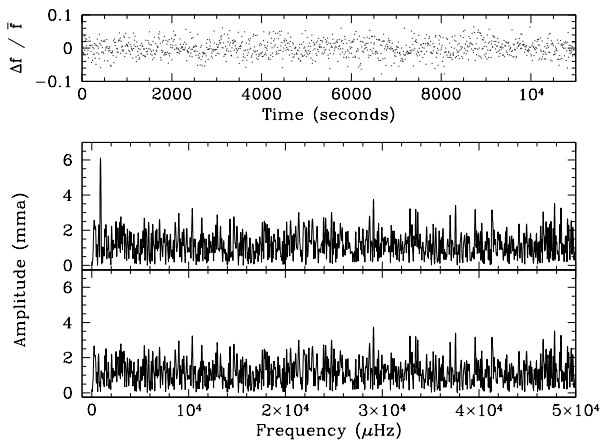}
\end{center}
\caption{Time-series photometry and analysis from the discovery run of
  \dqv.  \emph{Top panel:} Photometry from 10 s exposures, plotted in
  terms of fractional amplitude f.  No significant modulation is
  obvious.  \emph{Middle panel:} Discrete Fourier transform of the
  data in the top panel.  A significant signal with amplitude 6.2 mma
  is seen at 897.3 $\mu$Hz.  \emph{Bottom panel:} Discrete Fourier
  transform after pre-whitening the photometric data by the 897.3
  $\mu$Hz signal.  No other significant peaks are
  present.  \label{fig.disc}}
\end{figure}

A visual inspection of the
light curves reveals no obvious signs of variability in either raw or
smoothed light curves; an example light curve from the discovery run
is shown in the upper panel of Figure \ref{fig.disc}.  The measured standard deviation in the photometry of individual data points is 16.6 mma  (1
mma $= 10^{-3}$ fractional amplitude).

In the middle panel of Figure \ref{fig.disc} we show the discrete
Fourier transform (DFT) of the unsmoothed data from the discovery
(first) observing run.  A clear and significant signal is seen at a
frequency of 897.3 $\mu$Hz with an amplitude of $6.2\pm 1.0$ mma.  The DFT of the data after
pre-whitening by this single frequency reveals no other significant
signal (bottom panel).

Combining data from multiple observing runs can reduce the noise in
the DFT and permit more precise measurements of the frequency,
amplitude, and phase of the photometric modulation, \emph{if} the modulations
are coherent between each run.  Our preferred methodology involves
bootstrapping to determine a period precise enough to bridge multiple
runs without losing a cycle count
\citep[e.g.][]{1985ApJ...292..606W}.  Unfortunately, weather and scheduling frustrations resulted in 
multiple-month gaps between successful observing sequences that are too
large to bridge with this method.

\begin{figure*}
\begin{center}
\includegraphics[width=0.75\textwidth]{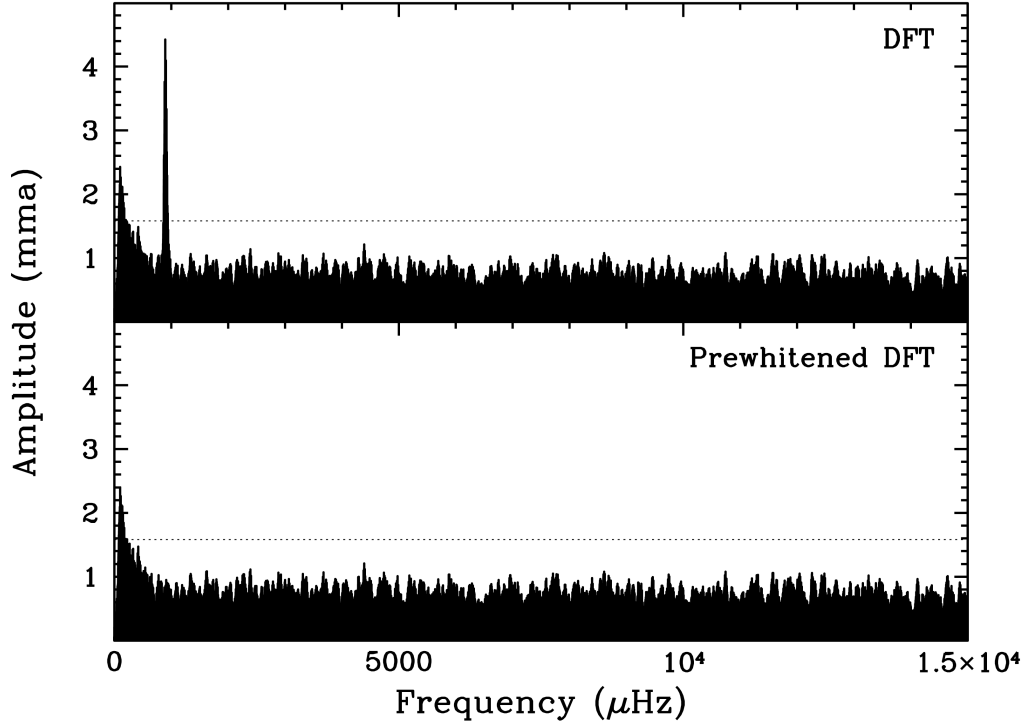}
\end{center}
\caption{ \emph{Top panel:} The discrete Fourier transform of the combination of all
  time-series photometry between 2008 December and 2010 April. \emph{Bottom panel:}
The discrete Fourier transform of the same combined data after pre-whitening by the single significant frequency.  In both panels, the horizontal dashed line indicates the 99\% false-alarm
  probability level.  Only one significant peak, the $P=1115.64751$ s
  modulation, is observed. \label{fig.dft_combo}} 
\end{figure*}

\begin{figure}
\begin{center}
\includegraphics[angle=270, width=0.49\textwidth]{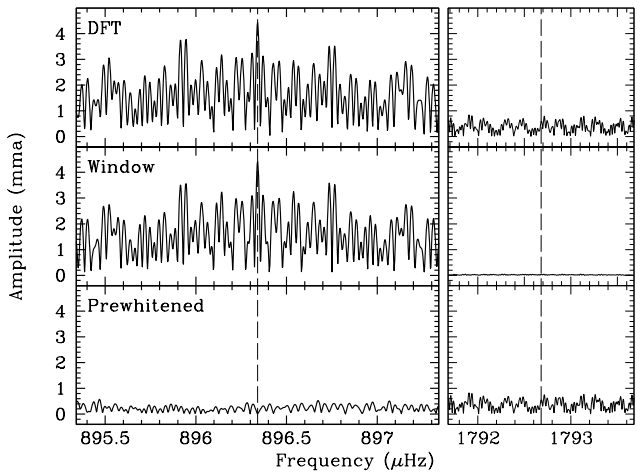}
\end{center}
\caption{\emph{Top panel:} The discrete Fourier transform (DFT) for the combined data from 2008 December through 2010 April.  \emph{Middle panel:} the DFT of the spectral window
  function for a single-frequency sinusoidal modulation with $P=1115.64751$ s. \emph{Bottom panel:} The DFT of the combined data set after pre-whitening by the observed modulation at $P=1115.64751$ s.  In each panel, the left portion highlights frequencies around the observed mode (indicated by the dashed line), and the right portion highlights frequency space surrounding the expected
  location of a first harmonic (dashed line).  Pre-whitening by a single mode is
  removes nearly all power from the DFT, and any harmonic must
  have an amplitude of $\lesssim 1$ mma.\label{fig.dft_combo_expand}}
\end{figure}

We therefore make the assumption that the photometric modulations are
coherent across our entire data set, and we calculate a DFT for the
entire combined data set; this DFT is shown in Figure
\ref{fig.dft_combo}.  Expanding the plot about the primary peak, we
see complex aliasing (Figure \ref{fig.dft_combo_expand}).
The highest peak, with frequency $f=896.340458(54) \,\mu{\rm Hz}$
[$P=1115.64751(67)$ s], is the only alias consistent with the
bootstrapped periods from the individual observing runs in 2008 December\footnote{We consider all data obtained between 2008 December 29 and 2009 January 4 as a single observing run},
2009 April, and 2010 January/February (i.e., the frequency of this alias is consistent with the frequency of the highest peak in each of the combined runs within calculated errors).  The window function for this
period is consistent with the observed aliasing, and pre-whitening the
light curve by this best-fit period removes all significant structure
from the DFT (Figure \ref{fig.dft_combo_expand}, middle and bottom
panels, respectively).  We therefore believe that this peak represents
the true frequency of modulation.

We determine the amplitude and phase of the modulation using
least-squares fitting of a sine wave to the combined data set.  The
resulting fit gives:
\begin{equation}
A(t) = A_0\sin[2\pi f (t-T_0)]\,,
\label{eqn.sine}
\end{equation}
where $A(t)$ is the observed fractional modulation, $A_0 = 4.42\pm
0.24$ mma is the amplitude of the modulation, $f$ is the frequency, $t$ is the BJD$_{\rm TDB}$ time of
observation \citep[Barycentric Julian Date of the Barycentric Dynamical Time, e.g.][]{2010PASP..122..935E}, and $T_0 = {\rm BJD}_{\rm TDB}\, 2454830.00459(11)$ d.

\begin{figure}
\begin{center}
\includegraphics[angle=270,width=0.49\textwidth]{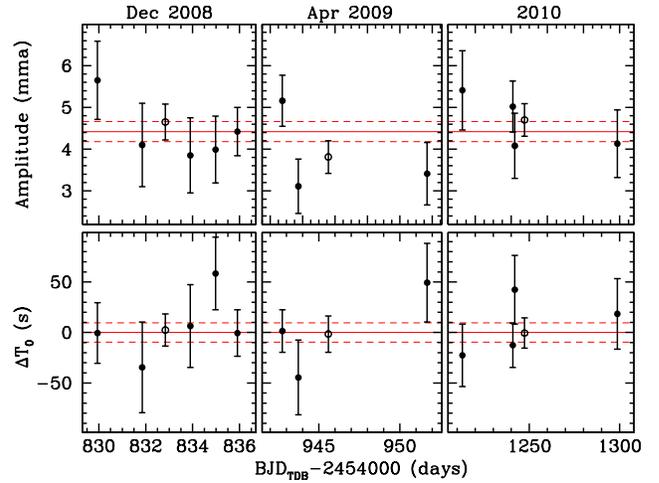}
\end{center}
\caption{Plots of the amplitude (top panels) and change in phase (equivalent to an $O-C$ diagram, bottom panels) of the $P=1115.64751$ s
  mode as a function of time.  Small filled points indicate
  individual runs; open circles indicate fits to the combined data set
  in each panel.  Horizontal lines are the best fit values (solid)
  and 1$\sigma$ errors (dashed) for the fit to the entire data
  set. No significant variability about the best-fit value is
  observed across the 15-month time span. The online version of the figure uses color for clarity.\label{fig.stability}}
\end{figure}

We also use the best-fit frequency to determine amplitudes and phases
for individual observing runs. These results, determined again via
least-squares fitting, are given in Table \ref{tab.indiv_fits} and
plotted in Figure \ref{fig.stability}.  There is no significant
variation of either the amplitude or $T_0$ across the 15 months,
indicating that the photometric modulations were coherent and of
constant amplitude over that time span.  If the photometric
modulations remain stable, it will be possible to constrain the rate
of change of the period ($\dot P$) and, potentially, discriminate
between various models of DQVs and their evolution.

\begin{deluxetable}{lcccc}
\tablecolumns{5}
\tablewidth{0pt}
\tablecaption{Fits for the 1115.64751 s mode\label{tab.indiv_fits}}
\tablehead{\colhead{Run(s)} & \colhead{$A$} &
\colhead{$\sigma A$} & \colhead{$T_0 - T_{0,all}$} & \colhead{$\sigma T_0$}
\\ & \colhead{(mma)} & \colhead{(mma)} & \colhead{(s)} & \colhead{(s)}}
\startdata
Entire Set & 4.42 & 0.24 & 0.0 & 9.5 \\
All 2010     & 4.70 & 0.39 & $-1$ & 15 \\
All 2009 Apr & 3.81 & 0.39 & $-2$ & 18 \\
All 2008 Dec & 4.65 & 0.43 & $2$ & 16 \\ 
A1807 & 5.65 & 0.94 &  $-1$ & 30 \\
A1811 & 4.10 & 1.00 & $-35$ & 45 \\
A1816 & 3.85 & 0.90 & $6$ & 41 \\
A1820 & 3.99 & 0.80 & $58$ & 36 \\
A1822 & 4.42 & 0.58 & $-1$ & 23 \\
A1858 & 5.16 & 0.61 & $1$ & 21 \\
A1861 & 3.11 & 0.65 & $-45$ & 37 \\ 
A1871 & 3.41 & 0.75 & $49$ & 39 \\
A2059 & 5.41 & 0.95 & $-23$ & 31 \\
A2076 & 5.02 & 0.61 & $-13$ & 22 \\
A2080 & 4.08 & 0.78 & $42$ & 34 \\
A2117 & 4.13 & 0.81 & $18$ & 35 \\
\enddata
\tablecomments{$T_0 - T_{0,all}$ is equivalent to the more familiar $O-C$ quantity.}
\end{deluxetable}

After pre-whitening of the light curve of \dqv by the best-fitting
solution above, we search for additional pulsational frequencies.  No
other obvious peaks are seen in the DFT.  We calculate false-alarm
probabilities using the method of Scargle (1982) as implemented by
Kepler (1993).  This method gives a 99\% false-alarm probability
amplitude of 1.54 mma; no peaks appear in the DFT higher than this
amplitude for frequencies $\gtrsim 500 \mu{\mathrm Hz}$. In
particular, no harmonic is observed at a frequency of 1793 $\mu$Hz
(right panels of Figure~\ref{fig.dft_combo_expand}).

\begin{figure}
\begin{center}
\includegraphics[width=0.48\textwidth,bb = 18 11 564 740,clip]{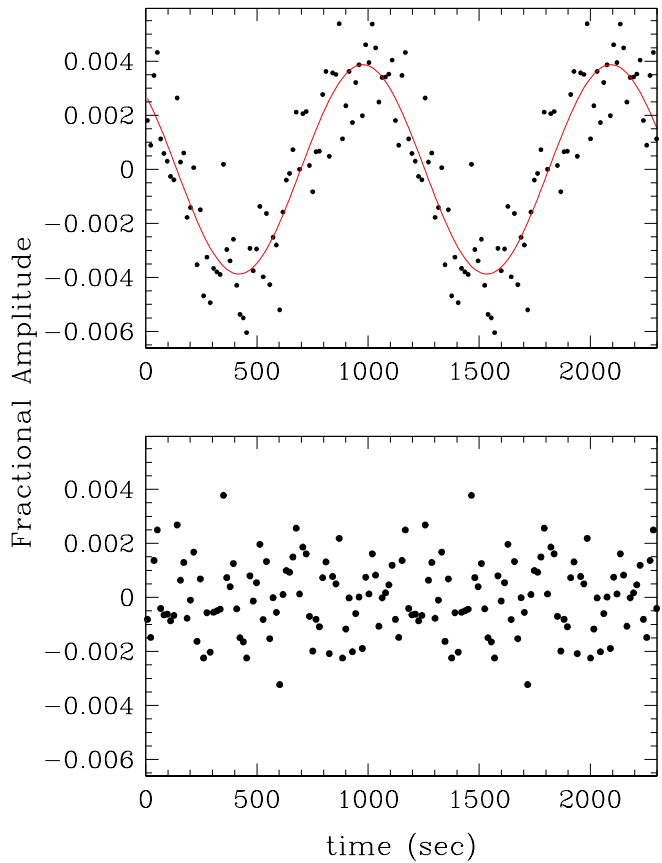}
\end{center}
\caption{\emph{Top panel}: The light curve of \dqv folded at its sole observed period
  of 1115.64751 s (points) and a sine wave of the same mean
  amplitude and period (solid curve, red in the online version of the figure).  Each point is the average of $\sim 60$ individual photometric measurements.  Two cycles are plotted for clarity.  \emph{Bottom panel}: The folded light curve after pre-whitning by the single observed period.  No residual signal is evident; the rms scatter in the residuals is 1.4 mma, consistent with the photometric measurement errors.   The residuals show no significant structure, in agreement with the lack of significant harmonics in the DFT. \label{fig.pulse}}
\end{figure}

As the light curve of \dqv is dominated by a single frequency of
modulation, we have computed a pulse shape by folding the data at the
1115.64751 s periodicity.  The individual measurements are averaged in 100 bins based on phase, giving approximately 60 photometric points per bin.  The resulting pulse shape is plotted in Figure
\ref{fig.pulse} along with a sine wave of the same mean
amplitude and period.  Within the scatter, the pulse shape looks
sinusoidal. 

The bottom panel of Figure \ref{fig.pulse} shows the folded light curve after pre-whitening the data by the single observed frequency.  No significant residual is seen, consistent with the interpretation of the pulse shape as a sine wave with no significant harmonics.   The root-mean square of the pre-whitened data is 1.4 mma, which is consistent with predictions for the scatter in each bin assuming typical photometric errors and Poisson statistics.

\section{Spectroscopic Observations and Analysis\label{sec.spec}}

The SDSS spectrum of \dqv is of moderate signal-to-noise ($\approx 28$
per resolution element at 4600\AA).  This was sufficient to identify
the atomic carbon lines and the Zeeman splitting observed by
\citet{2003AJ....126.2521L} and \citet{2003ApJ...595.1101S}, but not
to obtain precise \teff and \logg (see Figure 6).

In order to better constrain the atmospheric parameters of \dqv, we
obtained two sets of spectroscopic observations.  We obtained three exposures at the MMT on the night of UT 2007 Dec 15 totaling 3600 sec (3x1200) with the Blue Channel
Spectrograph. We used the 500 line mm$^{-1}$ grating with a 1\arcsec
~slit width, resulting in an $\sim 3.6$\AA~FWHM spectral resolution over a
wavelength range of $3400-6100$ \AA. These were reduced using standard IRAF routines and flux-calibrated using the standard stars \objectname{BD+284211} and \objectname{Feige 34}.

We obtained a second data set on
UT 2010 Feb 08 with the Low Resolution Imaging Spectrometer
\citep[LRIS,][]{1995PASP..107..375O} on the Keck I telescope.  We took
these observations in long-slit mode using the blue side of the
spectrograph with the atmospheric dispersion corrector.  We used the
600 line mm$^{-1}$ / 4000\AA ~grism in combination with the D560
dichroic and a slit width of 1\arcsec; the resulting data have a spectral
resolution of $\approx 4.1$\AA ~FWHM.  We obtained 2 exposures of 900
s each.

The Keck data were reduced using the \emph{onedspec} package in IRAF.
Overscan regions were used to subtract amplifier bias.  The internal
flat field was used to remove pixel-to-pixel sensitivity variations;
small amplitude ($\approx 0.5\%$), large-scale ($\sim 100$\AA)
residuals are present in the final flat-field.  Cosmic rays were
identified using the L.-A. Cosmic routine of
\citet{2001PASP..113.1420V}.  The flat-fielded exposures were
averaged; pixels flagged as cosmic rays were exluded from the
averaging.  

\begin{figure}
\begin{center}
\includegraphics[width=0.49\textwidth]{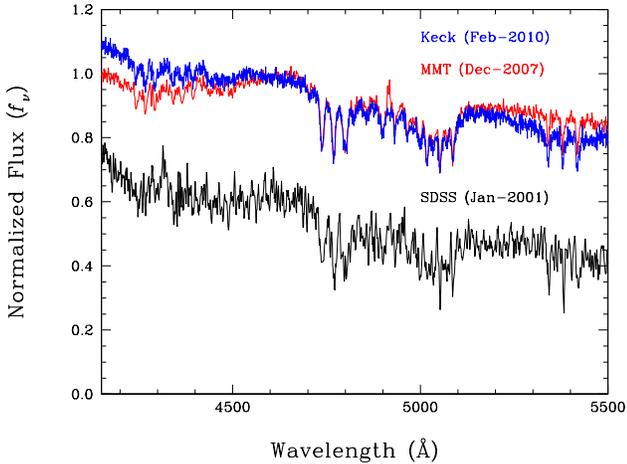}
\end{center}
\caption{Normalized spectroscopic data for SDSS J1036+6522. The small
  differences between the continuum level of MMT (red) and Keck (blue)
  data are due to errors in flux calibration. A three-point average
  smoothing window and a downward shift has been applied on the the
  SDSS data for clarity.  The majority of the visible features are
  \ion{C}{1} with magnetic splitting. \label{fig.spec}}
\end{figure}

We extracted the one-dimensional spectrum and applied wavelength
calibrations derived from arc lamp spectra.  We also derived and
applied an instrumental response calibration derived from the
spectrophotometric standard star \objectname{G 193--74} (spectral type DC); we make no
attempt at absolute spectrophotometry.  The resulting spectra, shown
in Figure \ref{fig.spec}, have similar signal-to-noise ratios of
$\approx 230$ per resolution element.

\subsection{Spectral Modeling}
One major difficulty in determining physical parameters of \dqv was
that tools for determining \teff and \logg for strongly magnetic,
carbon-helium mixed atmospheres simply did not exist.  Therefore,
one of us (PD) set out to develop model atmospheres appropriate for
analyzing this star.

\subsubsection{Line Splitting}
In the presence of a weak magnetic field, typically up to a few Tesla
(1 T = $10^4$ G), atoms can be described in the Zeeman regime. In that
regime, an external magnetic field can be treated as a small
perturbation to the Hamiltonian and the energy level properties can be
obtained from first order perturbation theory. Assuming that the atom
is described by L-S coupling (a decent approximation for light atoms),
J and $m_J$ are good quantum numbers. In the presence of a weak
magnetic field, an atomic level with total angular momentum J splits
into 2J+1 levels with magnetic quantum number $m_J$ = -J,$\ldots$,
+J. In that regime, the position of all magnetically shifted
transitions between an upper and a lower level can easily be computed
using the selection rule ($\Delta$m = 0, $\pm$1) and $\nu_{ul} = (E_u
- E_l)/h$, where $E_i = E_{i0} + (e\hbar/2m_e c) g_iBm_J$, B is the field
strength, $E_{i0}$ is the unperturbed energy of level {\it i} and
$g_i$ is the Lande factor of level {\it i} \citep[for recent
applications in WD stars, see][]{Farihi2011,Zuckerman2011}.

For stronger magnetic fields able to produce splitting on a J-level
that is comparable with the energy separation between the different
J-levels of a term, L and S decouple and J is no longer a good quantum
number. When this happens, the perturbation theory (Zeeman effect) can
no longer be applied. In this regime, called the Paschen-Back (PB)
effect regime, the energy levels have to be found by the
diagonalization of the total Hamiltonian $H$ = $H_0$ + $H_B$ where
$H_0$ is the unperturbed Hamiltonian and $H_B = e\hbar/2mc~(\mathbf{L}
+ 2\mathbf{S})\cdot\mathbf{B}$.

Although the Paschen-Back effect was successfully interpreted within
the framework of quantum mechanics long ago \citep[e.g. see classic
textbooks such as][]{Condon}, very few astronomical applications are
found in non-degenerate stars. The main reason is that fields found in
main sequence stars are not large enough to push most lines out of the
Zeeman regime (sunspot field strengths are typically between $1-3$ kG,
while there are cases of a few Ap stars with field up to 30 kG), so
only when the fine-structure levels of a term have abnormally small
separation must the partial Paschen-Back regime be taken into
account. For such weak fields (by WD standards) the effect was
only evidenced through the very small asymmetry that it induces in the
line profiles \citep{Mathys}. Line profile calculations including the PB
effect have been recently incorporated in stellar models only for a
few sensitive lines in Ap stars \citep[\ion{Fe}{2} multiplet 74 and Li
$\lambda$ 6707 doublet, see][]{Stift,Kochukhov} and the Sun
\citep[molecular lines and \ion{He}{1} $\lambda$10830,
see][]{Asensio,Berdyugina,Socas}.

Magnetic fields strong enough to push the lines into the PB regime are
commonly found in WD stars \citep[see the latest review
by][and references therein]{Jordan09}. Yet since the vast majority of
known magnetic WD stars only show hydrogen line splitting,
only hydrogen-rich magnetic model atmospheres have been developed in
detail to this date. However, the population of magnetic WDs
with heavy element features is growing rapidly thanks to the work on
DQ \citep[see][]{Schmidt95}, hot DQ \citep[][]{2009JPhCS.172a2012D}
and DZ WDs \citep[i.e., LHS~2534 and
G165-7][]{Reid01,Dufour06}. Accurate determination of the atmospheric
parameters for such stars requires the development of
a new generation of magnetic model atmospheres that include various
field geometries, radiative transfer for all Stokes components, as well
as splitting in the PB regime. Although such models are not available
yet, it is still possible to obtain acceptable results with a few
reasonable approximations.

\subsubsection{Models and Synthetic Spectra}

As mentioned above, detailed magnetic model atmospheres appropriate for
WD stars showing heavy elements are currently
unavailable. Nevertheless, it is possible to obtain a good first order
approximation with some reasonable assumptions. We first assume that
the thermodynamic structure is not affected by the magnetic field, a
conservative approximation for fields of a few MG according to
\citet{Jordan92}. We also calculate our synthetic spectrum using
normal radiative transfer, though unfortunately it is currently unknown to
what extent the use of formal radiative transfer equations for all
Stokes parameters would affect our results. Finally, we simply assume
a uniform magnetic field strength over the surface of the stars (which
seems to be indicated by our observational data, in particular the
sharp and well separated components of the magnetically split lines
in Figure 8). Future work, which should address all these issues,
will be presented in due time. Equipped with this simple theoretical
framework, we next calculate a grid of model atmospheres appropriate
for warm DQ stars.

Our LTE model atmosphere code is similar to that presented in
\citet{Dufour2005} and \citet{2008ApJ...683..978D} except that we use
absorption line data from the Vienna Atomic Line Database (VALD). Our
model grid covers a range of effective temperature from $\teff =
11,500-19,000$~K in steps of 500~K, $\logg = 8.0$ to 9.5 in steps of
0.5 dex, and from $\log (\mathrm{C}/\mathrm{He})= 0.0$ to $-3.5$ in
steps of 0.5 dex.

Detailed synthetic fluxes are then obtained from these atmospheric
structures by first including line splitting in the Zeeman regime for
a magnetic field of 3 MG. This value is obtained from measuring the
separation of the $\pi \, (\Delta m=0)$ and $\sigma\, (\Delta m=\pm 1)$ components of the well isolated
\ion{C}{1} 5380\AA ~triplet, shown in Figure \ref{fig.spec}.  Note that the separation
between the components for this transition are not different in the PB
regime.  To estimate the error, we also calculated a few supplementary model with magnetic field slightly different then 3 MG. From this experiment, we find that varying the field by more than 0.2 MG changes the line positions sufficiently to be detected by visual inspection. Hence, we adopt a field strength of $3.0 \pm 0.2$ MG.

\begin{figure}
\begin{center}
\includegraphics[scale=0.35,angle=-90]{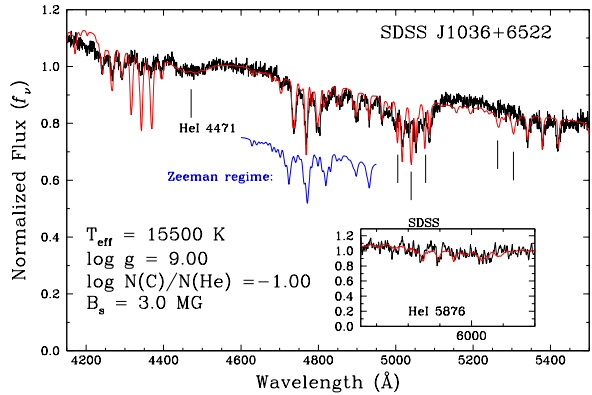}
\end{center}
\caption{Best fit of the Keck spectroscopic data. Insert shows SDSS
  coverage of the \ion{He}{1} $\lambda5876$ line (a three-point average smoothing
  window has been applied for clarity). Some of the strongest
  semi-forbidden transitions that were treated in the Zeeman regime
  (see text) are marked by ticks and clearly not well matched. Below these spectra is a synthetic spectrum with
  the same atmospheric parameters but calculated assuming splitting in
  the Zeeman regime; mismatches with the data are obvious. \label{fig.specfit}}
\end{figure}

While a 3 MG magnetic field is appropriate to reproduce the observed splitting in the 
$\sim 5380$\AA ~features, it fails to reproduce the
position of the carbon feature between $\sim 4700-5000$ \AA ~(see the
synthetic spectra calculated under the Zeeman regime assumption in
Figure \ref{fig.specfit}). As explained above, the most likely explanation is that the
magnetic field is sufficiently strong to push some of the lines in the
Paschen-Back regime where the position and strength of spectral
features differ to that predicted by simple Zeeman calculations.

\begin{figure}
\begin{center}
\includegraphics[scale=0.48]{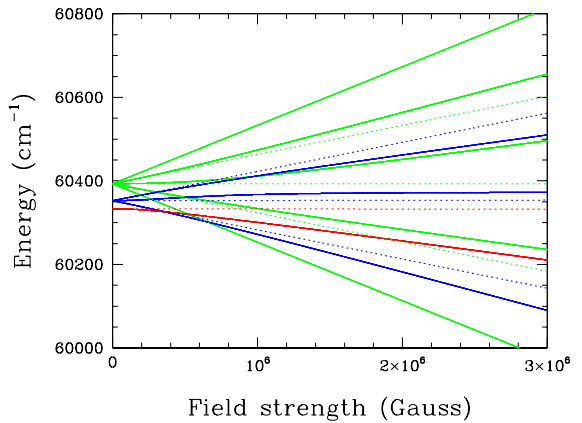}
\end{center}
\caption{Magnetic splittings of the 3 lower sub-levels of the $^3P$
  term of the \ion{C}{1} multiplet near 4775\AA. Dotted lines show the energy
  level as a function of magnetic field strength assuming Zeeman
  splitting while the full line is the result of the calculations in
  the Paschen-Back regime (see text). \label{fig.split}}
\end{figure}

Indeed, as can be observed in Figure \ref{fig.split}, the position of
the energy levels produced by a 3 MG magnetic field can sometimes be
significantly different in the Zeeman and Paschen-Back regime. Under
such circumstances, the energy levels and line strength values due to
the magnetic field perturbation must thus be obtained by the
diagonalization of a set of matrices for each of the transitions
included in our synthetic spectra. We thus proceeded in calculating
the line strength and positions of all the components of magnetically
split \ion{C}{1} lines in the PB regime \citep[the relevant equations are
described in detail in Section 3.4 of the monograph by][]{PISL2004}. NIST Atomic Spectra Database energy levels were
used as input for the PB calculations, though the line
strengths of each transition are obtained by re-normalizing
appropriately the sum of all the $\log g\!f$ of a given multiplet taken
from the VALD database. 

Our final grid of synthetic spectra is calculated with a mean magnetic
field of 3 MG based on the separation of the 5380\AA ~lines,
which are transitions between $^1P$ terms that are correctly described
with the Zeeman regime. Since the positions of the \ion{He}{1} lines
are correctly reproduced in the Zeeman regime, it was unnecessary to
perform the lengthy PB calculation for helium for this exploratory
analysis.

Finally, we note that our \ion{C}{1} line list contained
a few semi-forbidden transitions that violate the selection rules
($\Delta S = 0$ and/or $\Delta L = \pm1,0$). The PB formalism
described above cannot be used for these lines and a more general
approach needs to be implemented in that case. Unfortunately, the
relevant data necessary for such calculations are currently
unavailable. We thus simply modeled those transitions assuming normal
Zeeman splitting patterns, an approximation that does not correctly
represent the positions and strength of such lines (see, for example,
tick marks in Figure \ref{fig.specfit}).

\subsubsection{Fitting Technique, Results and Discussion}

We attempted to fit the spectra using the usual non-linear least
square routines commonly used in spectral fitting, but these led to
regions of parameter space that made little sense.  For example, a fit
might converge successfully on a minimum $\chi^2$, but visual
inspection would reveal obvious shortcomings such as not reproducing
the 4471\AA ~\ion{He}{1} dip, not matching the slope, or matching the strengths
and positions of one group of lines while another set was poorly
fit. It is likely that the models do not perfectly reproduce reality
because of the numerous approximations, and some badly reproduced
regions (such as the group of \ion{C}{1} lines near 4340\AA ~that are always
too strong, the semi-forbidden transitions, etc.) are simply pulling
the solution away from realistic values. Even when we fixed \logg and
\teff, the carbon abundance would wander to unrealistic values that
resulted in obvious visual mismatches.  We therefore resorted to
visual inspection for selecting the best model.

Another difficulty is the strong group of \ion{C}{1} lines near
4340\AA.  Choices of parameters that fit this complex well result in
very poor fits for the rest of the spectrum; there is no combination
of parameters in our model grid that can achieve a satisfactory fit
over all spectral region simultaneously. This could imply that our
numerous approximations affect the temperature structure in such a
way to poorly reproduce this complex or perhaps that the $\log g\!f$ values
for these lines are incorrect.

Our visual inspection therefore starts with the \ion{C}{2} lines near
4267\AA. Since this is an ionized feature, it provides decent
constraints on \teff.  We then use the broad \ion{He}{1} 4471\AA ~dip
as a main indicator of carbon abundances, as this dip proves quite
sensitive to $\log (\mathrm{C}/\mathrm{He})$.  In parallel, the MMT
and SDSS spectra include the \ion{He}{1} 5876\AA ~line, which we used
as a confirmation of the abundance.  This line also allows some of the
coolest solutions to be rejected.

Our best solution, shown in Figure \ref{fig.specfit}, has $\teff
=15500$ K, $\logg=9$, and $\log (\mathrm{C/He})=-1.0$. The solutions
are degenerate; similar quality solutions can be found by compensating a downward
change in \teff with downward changes in \logg and $\log
(\mathrm{C}/\mathrm{He})$.  We estimate our correlated uncertainties in these parameters as follows. A $\sim 1000$K change in our best solution's \teff resulted in a $\sim 0.5$dex change in $\log
(\mathrm{C}/\mathrm{He})$, with a lower \teff resulting in less carbon and vice-versa. A change in \logg has little effect on the carbon abundance as long as we stay close to $\logg=9.0$; at 15500K and $\logg=8.5$, $\log (\mathrm{C}/\mathrm{He})$ is only 0.2 dex higher. At $\logg=8.0$, the carbon abundance increases by about 0.7 dex as compared to $\logg=9$. Even though $\logg=9$ provides the most appealing solution visually, given all the approximations and the preliminary nature of our calculations, our value of $\logg$ should not be considered a precision measurement. Much more work is needed before we can confidently determine the mass of this object.

\begin{figure}
\begin{center}
\includegraphics[scale=0.35,angle=-90]{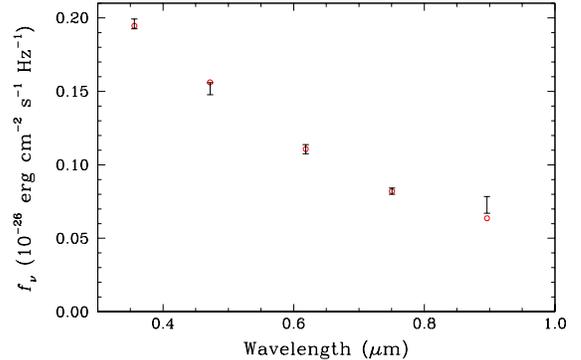}
\end{center}
\caption{SDSS photometric observations in the $ugriz$ bands (error bars)
  and average model fluxes for our favored atmospheric parameters (circles, red in the online version of the figure, see text). The photometric observations have been corrected
  for extinction as prescribed by the reddening maps of
  \citet{1998ApJ...500..525S}.\label{fig.photcomp}}
\end{figure}

Carbon-rich solutions, favored for hot DQs,
do not fit well.  However, we can impose some useful additional
constraints on these parameters. For instance, we compared the SDSS
photometry of \dqv with synthetic photometry calculated by folding the
model spectra through filter response curves. The photometric data
(Figure \ref{fig.photcomp}) seem to favor a solution in the
$15000-16000$ K range, a temperature range which is corroborated by the
\ion{He}{1} 5876\AA ~line, although the unknown reddening makes this
a bit uncertain. This conclusion also supports the qualitative best
matching model above.

\subsection{What is \dqv?}

Based on the spectroscopic and photometric analysis, \dqv seems to be
an intermediate object between the classical, cool DQ stars
\citep[$\teff\lesssim 10,000$ K, e.g.][]{Dufour2005} and hot DQ stars
\citep[$18,000\lesssim\teff\lesssim 24,000$
K,][]{2008ApJ...683..978D,2007Natur.450..522D}.  Its high carbon
abundance also seems to make it akin to the second sequence of
carbon-rich cool DQs \citep{Dufour2005,Koester2006} and two DQs with
similar temperature, \objectname{G35$-$26} \citep{1990ApJ...361..197T}
and \objectname{G227$-$5} \citep{1985ApJ...288..746W}.  In addition,
the high magnetic field of the star also appears to be a typical
characteristic of many of the hot DQs \citep{2009JPhCS.172a2012D}.

Although the high apparent surface gravity should be considered
preliminary, a growing body of evidence suggests that hot DQs and the
carbon-rich sequence of cool DQs are massive WDs.  This
evidence includes unpublished parallaxes of G35$-$26 and G227$-$5
\citep[][C.C.~Dahn, private communication]{2003AJ....126.2521L}; the
high mass of the carbon- and oxygen-rich post-AGB star H1504+65
\citep{2004A&A...421.1169W}; the presence of a hot DQ in the open star
cluster \objectname{Messier 35}, in which the DA cluster member WDs
are all massive \citep{2006ApJ...643L.127W,2009ApJ...693..355W}; and
improved spectroscopic observations and model atmospheres of hot DQs,
which favor high surface gravities (Dufour et al., in prep).

We note however that these calculations are done with classical
approximations for the Stark broadening of \ion{C}{1} lines, so the
high gravity solution should be considered highly
uncertain. State-of-the-art Stark broadening parameters for \ion{C}{1} similar
to that calculated for \ion{C}{2} in \citet{2011arXiv1107.3066D} are
unfortunately not available yet.

In summary, \dqv is most probably a cooler version of \object{SDSS
  J1402+3818} and \object{SDSS J0005-1002}
\citep[see][]{2009JPhCS.172a2012D} and/or a hotter version of G227$-$5
and G35$-$26. \dqv could thus be a transition object between the
coolest carbon-rich hot DQs and the hottest, cool helium-rich DQ white
dwarfs.  There is thus growing evidence that this class of WDs indeed
forms a distinct and fairly continuous evolutionary sequence of
massive hydrogen- and helium-deficient stars, as proposed by
\citet{2008ApJ...683..978D}.
Further analysis on this type of object will require improved models
including various magnetic field geometries, solving the radiative
transfer equation for all Stokes parameters (spectropolarimetric
observations would be especially useful to test the eventual
predictions from such models), and inclusion of state-of-the-art Stark
broadening parameters for \ion{C}{1} once available.

\section{Discussion\label{sec.discuss}}

The detection of variability in \dqv increases yet again the number of
known DQVs.  However, the spectroscopic analysis raises the question of
whether or not \dqv belongs in the same class as the other known DQVs.

There are some crucial differences between \dqv and the hot DQVs, first and foremost being \teff, with \dqv being significantly cooler than the other DQVs, as discussed above.
Second, the atmospheric composition of \dqv is different than both the hot and cool DQs.  Hot DQs are carbon-dominated, with $\log({\mathrm C}/{\rm He})\geq 0$, and He usually undetected \citep{2008ApJ...683..978D}.  In the ``second-sequence" cool DQs,   $\log({\mathrm C}/{\rm He})\lesssim -3$ \citep{Koester2006,Dufour2005}.  Again, \dqv is intermediate, with $\log({\mathrm C}/{\rm He})\approx -1$. This observation furthers the hypothesis that \dqv is a transition object between hot and cool DQs.

We now explore the interpretations and implications of this conclusion.

\subsection{What is the cause of  the variability?} 

From an observational standpoint, the photometric variability of \dqv
is very similar to the other known DQVs.  These similar
characteristics include the period and amplitude of the photometric
variability, the dominance of a single frequency in the discrete
Fourier transform, and the stability of the frequency and phase.

Based on the time-series photometry, \dqv looks and acts like the
other members of the DQV class of objects.  Naively combining this
information with the spectroscopic evidence that this star belongs to
the same evolutionary sequence and object class as the hot DQVs, we
would come to the conclusion that \dqv is a typical DQV and is
variable for the same reason as the others, whatever that reason may
be.  

The difficulties arise when we attempt to explain the origin of the
photometric variations.  As described in the introduction, the
preferred explanation for the variability in DQVs has been nonradial
pulsations.  Even though there are differing physical mechanisms
and different predicted instability strips for pulsations in DQVs,
\dqv is far cooler than \emph{any} of these predictions (though we note that these models do not include a significant magnetic field).  Moreover, if our
atmospheric analysis is correct, the atmosphere of \dqv is helium
dominated and $\sim 10^4$ K cooler than the coolest observed
DBV \citep[e.g.][]{2009ApJ...690..560N}.

\citet{2008ApJ...678L..51M} hypothesize that the periodic
photometric variability could be due to an accreting system akin to
high-accretion states of \objectname{AM CVn} systems.  This speculation was based
solely on the observed pulse shape of SDSS J1426+5752.  This
hypothesis now appears highly unlikely.  High signal-to-noise spectra
of DQVs show no signs of accretion
\citep[e.g.][]{2009JPhCS.172a2012D,Dufour2011}, radial-velocity
variations have not been detected with cataclysmic-variable-like
periods \citep[][although orbital periods commensurate with the
photometric period have not yet been ruled out]{2009ApJ...702.1593G},
and no low-accretion state systems analogues have been identified.
Given that \dqv has a sinusoidal pulse shape, we find no evidence that
\dqv is an interacting binary system.

Another possibility, discussed by \citet{2008ApJ...683L.167D}, is that
this star is a WD analogue of rapidly oscillating Ap (roAp)
stars \citep[e.g.][]{1982MNRAS.200..807K}.  Given the presence of a
strong magnetic field, this oblique pulsator model seems to be a
distinct possibility.

It is also plausible that \dqv is a rapidly rotating WD, with the photometric variations due to the presence of a persistent spot, although several arguments could be made against this interpretation.  Accreting magnetic objects such as AM Her systems show photometric modulation due in part to the rotation of the WD  carrying the accretion column in and out of view, but \dqv shows no evidence of accretion.  In addition, the measured rotation periods of single magnetic WDs are on the order of hours and days to centuries \citep{1991ApJ...366..270S}, significantly longer than the photometric period of \dqv.  As the measured rotation rates of other classes of isolated pulsating WDs are also on the order of hours to days \citep[e.g.][and references therein]{2008ARA&A..46..157W,2008PASP..120.1043F}, the rotation hypothesis would require \dqv to be an outlier among single WDs.  Time-resolved spectropolarimetry may be able to determine if \dqv is rapidly rotating.

At present, it is unclear what the cause of the photometric
variability in \dqv is, or even if the variability has the same cause
as for the hot DQVs.

\subsection{Magnetic fields and pulse shapes}

Typically, large amplitude DAV and DBV pulsators tend to have relatively non-linear (non-sinusoidal) pulse shapes, while low-amplitude pulsators tend to be more sinusoidal.  This relation does not appear to hold among the known DQVs with sufficient data to calculate pulse shapes.  In the prototype DQV SDSS
J1426$+$5752, the amplitude of the first harmonic (6.7 mma)  is less than half that of the
fundamental \citep[17.5 mma;][]{2008ApJ...678L..51M}, while in the much lower amplitude DQV SDSS J2200$-$0741, the first harmonic is similar
in amplitude to the fundamental frequency at $\approx 7$ mma \citep{2008ApJ...688L..95B,2009ApJ...703..240D}.  These significant
harmonics and their phases result in decidedly non-sinusoidal, boxy
pulse shapes with long brightness maxima and short, narrow minima, the
opposite to the pulse shapes found in the high amplitude DAVs and DBVs \citep{2008ApJ...678L..51M}.  However, the similar-amplitude DQV SDSS J2348$-$0943 \citep[$\approx 8$ mma;][]{2008ApJ...688L..95B,2009ApJ...703..240D} is sinusoidal, and the low-amplitude DQV SDSS 1337$-$0026 \citep[$\approx 4$ mma][]{Dunlap2010} has a very non-sinusoidal pulse shape.  

Further, the three DQVs with non-sinusoidal pulse shapes all have detectable magnetic fields, while the sinusoidal SDSS J2348$-$0943 does not.  This has led to the suggestion by \citet{2009ApJ...702.1593G} that the magnetic field may be responsible for the non-linearities observed in many DQVs.

The folded pulse shape of \dqv appears to be sinusoidal.  This
qualitative statement is reinforced quantitatively by the lack of any
observed harmonics in the DFT.  
As
\dqv is strongly magnetic yet has a sinusoidal pulse shape (within the
limits of existing data), it is clear that strong magnetic fields in
DQVs do not always cause boxy pulse shapes.  Strong magnetic
fields in SDSS J1337$-$0026, SDSS J1426$+$5752 and SDSS J2200$-$0741 may well be the
cause of their unique pulse shapes, but \dqv shows that a sinusoidal
pulse shape in a DQV does not indicate the absence of a strong
magnetic field, at least assuming that the mechanism behind the
variability is the same for \dqv and the hotter DQVs.

\subsection{Non-variable hot DQs}

\citet{2008ApJ...678L..51M} claim that variability was not detected in several hot DQs, as well as that those stars
would not be expected to vary given their \teff and \logg and given
predictions for an instability strip.  In light of the low
amplitude of variability in \dqv and SDSS J1337-0026 \citep{Dunlap2010} and the typical detection threshold of $\sim
5-10$ mma in the data obtained for \citet{2008ApJ...678L..51M}, we feel compelled to mention that many of
the stars not observed to vary could in fact be variable at the same
level as \dqv.  Photometric precision of equal to or better than 1 mma would be required to rule out \dqv-like variability.

Therefore, we feel that the  claims of non-detection of variability in
\citet{2008ApJ...678L..51M} should be viewed with the caveat that these limits are not very significant. As \citet{Dufour2011} detect UV variability in four out of five target hot DQs, it could be that most or all hot DQs are in fact DQVs.  A re-analysis of the
time-series data for stars not observed to vary is underway, and more stringent limits on non-detections of
variability will be presented in a forthcoming paper.  These limits will also be crucial
for testing competing models for these stars, such as the different
proposed instability strips discussed in the introduction.

\section{Conclusions}
In summary, we present spectroscopy and time-series photometry of the white dwarf \dqv.  Our analyses lead us to the following conclusions:
\begin{itemize}
\item \dqv is photometrically variable with a period of 1115.64751(67) s and an amplitude of $0.442\% \pm 0.024\%$.  The period, amplitude, and phase of the modulations are constant within measurement errors over at least a 16 month interval.
\item \dqv has an effective temperature and carbon abundance located between the hot and cool DQs, suggesting that it is a transition object between those two populations of white dwarfs.
\item The star is magnetic, with a field strength of $3.0 \pm 0.2$ MG.
\item The lack of observed harmonics suggest that the photometric modulation is nearly sinusoidal.  If the cause of the photometric modulation is the same mechanism as in the hot DQVs, then the absence or strength of the star's magnetic field is not necessarily correlated with the pulse shape.
\item The mechanism of the photometric modulation in \dqv is unclear.  There is no evidence for accretion, the effective temperature is significantly cooler than pulsation models predict, and the photometric period is significantly shorter than the typical isolated white dwarf rotational period.
\item Photometric variability is common among hot DQ white dwarfs, but the low optical amplitudes of many known DQVs imply that non-detections of variability in any hot DQ are only significant if photometric precision is high ($\sim 0.1\%$).
\end{itemize}

 $\phantom{0}$
 
\acknowledgements K.A.W.~ is grateful for the financial support of
National Science Foundation award AST-0602288.  We would like to thank
E. Landi Delgl'Innocenti for useful discussions and his help in
modeling lines in the Paschen-Back regime. P.D is a CRAQ postdoctoral
fellow. We acknowledge the use of the VALD Database
\citep{Piskunov,Ryabchikova,Kupka99,Kupka00}. D.E.W, M.H.M., J.J.H, R.E.F., and K.I.W. acknowledge the support of the NSF under grant AST-0909107 and the Norman Hackerman Advanced Research Program under grant 003658-0252-2009.  We are very grateful for the technical support provided by D.~Chandler during the photometric observations and for the many fine improvements to the Struve telescope and telescope control spearheaded by J. Kuehne and the McDonald Observatory staff.  Ron Leck and R.~E.~Nather were indispensable in upgrading the instrumental software during these runs. We also appreciate the comments and insight of the anonymous referee.

{\it Facilities:} \facility{Struve (Argos)}, \facility{Keck:I (LRIS)},
\facility{MMT (Blue Channel)}

\end{document}